\documentstyle[12pt]{article}
\advance \topmargin by -\headheight
\advance \topmargin by -\headsep

\evensidemargin \oddsidemargin
\begin{document}

\topmargin -.6in

\def\rf#1{(\ref{eq:#1})}
\def\lab#1{\label{eq:#1}}
\def\br{\begin{eqnarray}}
\def\er{\end{eqnarray}}
\def\be{\begin{equation}}
\def\ee{\end{equation}}
\def\nn{\nonumber}
\def\ct{\cite}
\def\lb{\lbrack}
\def\rb{\rbrack}
\def\({\left(}
\def\){\right)}
\def\v{\vert}
\def\bv{\bigm\vert}
\def\lskip{\vskip\baselineskip\vskip-\parskip\noindent}
\relax
\newcommand{\nit}{\noindent}
\newcommand{\bi}[1]{\bibitem{#1}}
%
%
\def\PRL#1#2#3{{\sl Phys. Rev. Lett.} {\bf#1} (#2) #3}
\def\NPB#1#2#3{{\sl Nucl. Phys.} {\bf B#1} (#2) #3}
\def\NPBFS#1#2#3#4{{\sl Nucl. Phys.} {\bf B#2} [FS#1] (#3) #4}
\def\CMP#1#2#3{{\sl Commun. Math. Phys.} {\bf #1} (#2) #3}
\def\PRD#1#2#3{{\sl Phys. Rev.} {\bf D#1} (#2) #3}
\def\PRv#1#2#3{{\sl Phys. Rev.} {\bf #1} (#2) #3}
\def\PLA#1#2#3{{\sl Phys. Lett.} {\bf #1A} (#2) #3}
\def\PLB#1#2#3{{\sl Phys. Lett.} {\bf #1B} (#2) #3}
\def\JMP#1#2#3{{\sl J. Math. Phys.} {\bf #1} (#2) #3}
\def\PTP#1#2#3{{\sl Prog. Theor. Phys.} {\bf #1} (#2) #3}
\def\SPTP#1#2#3{{\sl Suppl. Prog. Theor. Phys.} {\bf #1} (#2) #3}
\def\AoP#1#2#3{{\sl Ann. of Phys.} {\bf #1} (#2) #3}
\def\PNAS#1#2#3{{\sl Proc. Natl. Acad. Sci. USA} {\bf #1} (#2) #3}
\def\RMP#1#2#3{{\sl Rev. Mod. Phys.} {\bf #1} (#2) #3}
\def\PR#1#2#3{{\sl Phys. Reports} {\bf #1} (#2) #3}
\def\AoM#1#2#3{{\sl Ann. of Math.} {\bf #1} (#2) #3}
\def\UMN#1#2#3{{\sl Usp. Mat. Nauk} {\bf #1} (#2) #3}
\def\FAP#1#2#3{{\sl Funkt. Anal. Prilozheniya} {\bf #1} (#2) #3}
\def\FAaIA#1#2#3{{\sl Functional Analysis and Its Application} {\bf #1} (#2)
#3}
\def\BAMS#1#2#3{{\sl Bull. Am. Math. Soc.} {\bf #1} (#2) #3}
\def\TAMS#1#2#3{{\sl Trans. Am. Math. Soc.} {\bf #1} (#2) #3}
\def\InvM#1#2#3{{\sl Invent. Math.} {\bf #1} (#2) #3}
\def\LMP#1#2#3{{\sl Letters in Math. Phys.} {\bf #1} (#2) #3}
\def\IJMPA#1#2#3{{\sl Int. J. Mod. Phys.} {\bf A#1} (#2) #3}
\def\IJMPB#1#2#3{{\sl Int. J. Mod. Phys.} {\bf B#1} (#2) #3}
\def\AdM#1#2#3{{\sl Advances in Math.} {\bf #1} (#2) #3}
\def\RMaP#1#2#3{{\sl Reports on Math. Phys.} {\bf #1} (#2) #3}
\def\IJM#1#2#3{{\sl Ill. J. Math.} {\bf #1} (#2) #3}
\def\APP#1#2#3{{\sl Acta Phys. Polon.} {\bf #1} (#2) #3}
\def\TMP#1#2#3{{\sl Theor. Mat. Phys.} {\bf #1} (#2) #3}
\def\JPA#1#2#3{{\sl J. Physics} {\bf A#1} (#2) #3}
\def\JSM#1#2#3{{\sl J. Soviet Math.} {\bf #1} (#2) #3}
\def\MPLA#1#2#3{{\sl Mod. Phys. Lett.} {\bf A#1} (#2) #3}
\def\JETP#1#2#3{{\sl Sov. Phys. JETP} {\bf #1} (#2) #3}
\def\CAG#1#2#3{{\sl  Commun. Anal\&Geometry} {\bf #1} (#2) #3}
\def\JETPL#1#2#3{{\sl  Sov. Phys. JETP Lett.} {\bf #1} (#2) #3}
\def\PHSA#1#2#3{{\sl Physica} {\bf A#1} (#2) #3}
\def\PHSD#1#2#3{{\sl Physica} {\bf D#1} (#2) #3}
\def\PJA#1#2#3{{\sl Proc. Japan. Acad.} {\bf #1A} (#2) #3}
\def\JPSJ#1#2#3{{\sl J. Phys. Soc. Japan} {\bf #1} (#2) #3}
\def\SJPN#1#2#3{{\sl Sov. J. Part. Nucl.} {\bf #1} (#2) #3}
%
\def\a{\alpha}
\def\b{\beta}
\def\c{\chi   }
\def\ca{{\cal A}}
\def\cm{{\cal M}}
\def\cn{{\cal N}}
\def\cf{{\cal F}}
\def\d{\delta}
\def\D{\Delta}
\def\eps{\epsilon}
\def\g{\gamma}
\def\G{\Gamma}
\def\vp{\varphi}
\def\grad{\nabla}
\def\h{ {1\over 2}  }
\def\hc{\hat{c}}
\def\hd{\hat{d}}
\def\hg{\hat{g}}
\def\hp{ {+{1\over 2}}  }
\def\hm{ {-{1\over 2}}  }
\def\k{\kappa}
\def\l{\lambda}
\def\L{\Lambda}
\def\lg{\langle}
\def\m{\mu}
\def\n{\nu}
\def\o{\over}
\def\O{\Omega}
\def\p{\phi}
\def\pa{\partial}
\def\pr{\prime}
\def\ra{\rightarrow}
\def\rh{\rho}
\def\rg{\rangle}
\def\s{\sigma}
\def\t{\tau}
\def\th{\theta}
\def\ti{\tilde}
\def\wti{\widetilde}
\def\inte{\int dx }
\def\xb{\bar{x}}
\def\yb{\bar{y}}
\newcommand{\eq}{=}
\def\tr{\mathop{\rm tr}}
\def\Tr{\mathop{\rm Tr}}
\def\partder#1#2{{\partial #1\over\partial #2}}
\def\ds{{\cal D}_s}
\def\wtwo{{\wti W}_2}
\def\lie{{\cal G}}
\def\alie{{\widehat \lie}}
\def\dlie{{\cal G}^{\ast}}
\def\elie{{\widetilde \lie}}
\def\edlie{{\elie}^{\ast}}
\def\hlie{{\cal H}}
\def\wlie{{\widetilde \lie}}
\def\rlx{\relax\leavevmode}
\def\inbar{\vrule height1.5ex width.4pt depth0pt}
\def\IZ{\rlx\hbox{\sf Z\kern-.4em Z}}
\def\IR{\rlx\hbox{\rm I\kern-.18em R}}
\def\IC{\rlx\hbox{\,$\inbar\kern-.3em{\rm C}$}}
\def\one{\hbox{{1}\kern-.25em\hbox{l}}}

\begin{titlepage}

June, 1999 \hfill{}
\vskip .6in

\begin{center}
{\large {\bf The Faddeev-Jackiw Approach and the Conformal Affine $sl(2)$ Toda Model Coupled to Matter Field}}
\end{center}

\normalsize
\vskip .4in

\begin{center}

{H. Blas{\begingroup\def\thefootnote{a}\footnote{e-mail: blas@ift.unesp.br}\addtocounter{footnote}{-1}\endgroup} and B.M. Pimentel{\begingroup\def\thefootnote{b}\footnote{e-mail: pimentel@ift.unesp.br}\addtocounter{footnote}{-1} \endgroup}}
\par \vskip .1in \noindent
Instituto de F\'{\i}sica Te\'{o}rica-UNESP\\
Rua Pamplona 145\\
01405-900 S\~{a}o Paulo, Brazil
\par \vskip .3in

\end{center}

\begin{center}
{\large {\bf ABSTRACT}}\\
\end{center}
\par \vskip .3in \noindent

The conformal affine $sl(2)$ Toda model coupled to matter field is treated as a constrained system in the context of Faddeev-Jackiw and the (constrained) symplectic schemes. We recover from this theory either, the sine-Gordon or the massive Thirring model, through a process of Hamiltonian reduction, considering the equivalence of the Noether and topological currrents as a constraint and gauge fixing the conformal symmetry.  
\end{titlepage}

\section{Introduction}
The Dirac method \ct{dirac} has been a standard method to deal with constrained systems for a long time. By using his approach one can obtain the Dirac brackets which are the bridges to the commutators in quantum theory. A few years ago, Faddeev and Jackiw (FJ)\ct{fj,ja} proposed a method of symplectic quantization of constrained systems for a first order Lagrangian, which appeared as an alternative to the traditional and succesful Dirac's approach. In the FJ method, the classification of constraints as first or second class, primary or secondary, is not necessary. It works, roughly, by solving the constraints and reducing the phase space of the system to the independent degrees of freedom. In \ct{cron} a constructive procedure for obtaining the canonical variables was proposed. An interesting variation inside the FJ method, known as ``symplectic quantization'' is due to Wotzasek, Montani and Barcelos-Neto \ct{wot,mont1,mont2,barc1}. Following the spirit of Dirac's work, this proposal works by imposing the stability of the constraints under time evolution. So, constraints are not solved but embedded in an extended phase space. This is a more suitable approach when some relevant symmetries must be preserved \ct{wot}.

In addition, integrable theories in two-dimensions have been an extraordinary laboratory for the understanding of basic nonperturbative aspects of physical theories. In the present paper we study the recently proposed conformal affine $sl(2)$ Toda model coupled to (Dirac) matter field (CATM) which is an example of a wide class of integrable theories presented in \ct{matter}. The zero curvature representation, the construction of the general solution and many other properties are discussed in \ct{matter}. This model possesses a Noether current depending only on the matter field and under some circunstances, it is possible to choose one solution in each orbit of the conformal group such that, for these solutions, the $U(1)$ current is equal to a topological current depending only on the Toda field. Such equivalence leads, at the classical level, to the localization of the matter field inside the soliton, and an additional feature; the masses of solitons and particles are proportional to the $U(1)$ Noether charge. These facts indicate the existence of a sort of duality in these models involving solitons and particles \ct{montonen}.

Here we carry out the classical analysis of the theory, once, the conformal invariance is gauge fixed by suitable choosing a classical vaccum for an auxiliary field, thus defining a related model which we call, affine Toda model coupled to matter field (ATM). Then, we  impose  the equivalence of the Noether and topological currents as a constraint which is incorporated to the action by means of Lagrange multipliers. It is interesting to see how the FJ and the constrained symplectic methods work in the analysis of the phase space of this submodel. Using these methods we show that this submodel turns out to be a parent Lagrangian \ct{hje} from which the sine-Gordon and the massive Thirring models are derivable. We thus show that there are (at least classically) two equivalent descriptions of this submodel, using either the Dirac or the Toda field. It will be also clear the duality exchange of the coupling regimes: $g \rightarrow 1/g$.
             
The paper is organized as follows. In section 2 we make a brief review of both the FJ and (constrained) symplectic formalisms. In section 3 we present some relevant aspects of the conformal affine $sl(2)$ Toda model coupled to matter (Dirac) field. Section 4 deals with the gauge fixed version of the model, the affine Toda model coupled to matter, in the original FJ \ct{ja} framework; the constraint is solved and Darboux's transformations are used to ``diagonalize'' the canonical sector of the theory. This reduction process results in the massive Thirring model. In section 5 we attack the same problem from the point of view of symplectic quantization \ct{wot,mont1,mont2,barc1}; where the constraints are not solved but considered as strong relations into the symplectic potential, and, simultaneously, incorporated to the canonical part by using Lagrange multipliers which are velocities. The outcome of this analysis is the Poisson brackets of either, the massive Thirring model or the sine-Gordon system, derived by gauge fixing the symmetries of the model in two different ways.

\section{The Faddeev-Jackiw formalism}

The Faddeev-Jackiw (F-J) approach \ct{fj,ja} avoids the separation of the constraints into first and second class and gives us a straightforward way to deal with constraint systems. A brief summary of this method is given below. In this section we use finite degree of freedom. The extension to the infinite degree of freedom case can be done in a straightforward way. Let us start with a first order in time derivative Lagrangian which may arise from a conventional second order one after the introduction of auxiliary fields. The general form of such a Lagrangian is 
  
\be
\lab{lag1}
L=a_{i}(\xi)\dot{\xi}^{i}-V(\xi).
\ee

Where the coordinates $\xi_{i}$, with $i=1,...,N$, stand for the generalized coordinates. Notice that when a Hamiltonian is defined by the usual Legendre transformations, V may be identified with the Hamiltonian H. 
   
This first order system is characterized by a closed two-form. If the two-form is not degenerated, it defines a symplectic structure on the phase space M, described by the coordinates $\xi_{i}$. On the other hand, if the two-form is singular, with constant rank on M, it is called a presymplectic two-form \ct{arno}. Thus, in terms of components, the (pre)symplectic form is defined by

\be
\lab{form}
f_{ij}(\xi)\eq\frac{\pa}{\pa{\xi}^{i}}a_{j}(\xi)-\frac{\pa}{\pa{\xi}^{j}}a_{i}(\xi)
\ee 
with the vector potential $a_{i}(\xi)$ being an arbitrary function of $\xi^{i}$. The Euler-Lagrange equations are given by   

\be
\lab{eqm1}
f_{ij}\dot{\xi}^{j}\eq\frac{\pa}{\pa{{\xi}}^{i}}V({\xi}). 
\ee

In the non-singular, unconstrained case the anti-symmetric $N{\mbox x}N$ matrix $f_{ij}$ has the matrix inverse $f^{ij}$, then $N=2n$, and \rf{eqm1} implies

\be
\lab{eqm2}
\dot{\xi}^{i}\eq f^{ij}\frac{\pa}{\pa{{\xi}}^{j}}V({\xi})
\ee
and the bracket will be defined by 
\be
\{\xi^{i}\;,\; \xi^{j}\}\eq f^{ij}
\ee

In the case that the Lagrangian \rf{lag1} describes a constrained system, the matrix $f^{ij}$ is singular which means that there is a set of relations between the velocities reducing the degrees of freedom of the system. Let us suppose that the rank of $f$ is 2n, so there exist $N-2n=N^{\prime}$ zero modes ${\bf v}^{\a}$, $\a\eq1,...,N^{\prime}$. The system is then constrained by $N^{\prime}$ equations in which no time-derivatives appear. Then there will be constraints that reduce the number of degres of freedom of the theory. Multiplying \rf{eqm1} by the (left) zero-modes ${\bf v}^{\a}_{i}$ of $f_{ij}$ we get   
\be
{\bf v}_{i}^{\a}f_{ij}\dot{\xi}_{j}\eq{\bf v}_{i}^{\a}\frac{\pa V(\xi)}{\pa {\xi}_{i}}\eq0.
\ee
These (symplectic) constraints appear as algebraic relations
\be
\lab{const}
\O_{\a}\equiv{\bf v}_{i}^{\a}\frac{\pa V(\xi)}{\pa {\xi}_{i}}\eq0.
\ee

By using Darboux's theorem \ct{ja,arno} one can show that an arbitrary vector potential, $a_{i}$, whose associated field strength $f_{ij}$ is non-singular, can be mapped by a coordinate transformation onto a potential of the form $a_{i}(\xi)\eq{1\o 2}\xi^{j}w_{ji}$ with $w_{ji}$ a constant and non-singular matrix. Then, the Darboux construction may still be carried out for the non-singular projection of $f_{ij}$ given in \rf{form}. Then the Lagrangian becomes
\be
\lab{lagz}
L\eq{1\o 2}\xi^{i}w_{ij}\dot{\xi}^{j}-V(\xi,z)
\ee
where $z$ denote the $N-2n$ coordinates that are left unchanged. Some of the $z^{\prime}s$ may appear non-linearly and some linearly in \rf{lagz}. Then using the Euler-Lagrange equation for these coordinates we can solve for as many $z^{\prime}s$ as possible in term of ${\xi}^{\prime}$s and other $z^{\prime}s$ and replace back in $V(\xi,z)$ so finally we are left only with linearly occuring $z^{\prime}s$. So, we can write the Lagrangian in the form   
\be
\lab{lagl}
L={1\o 2}\xi^{i}w_{ij}\dot{\xi}^{j}-V(\xi)-\l_{k}\Phi^{k}(\xi),
\ee
where we have renamed the linearly occuring $z's$ as $\l_{k}$. We see that these $\l_{k}$ become the Lagrange multipliers and $\Phi^{k}(\xi)$ are the constraints. To incorporate the constraints we solve the equations 
\be
\Phi^{k}(\xi)\eq0
\ee
and replace back in \rf{lagl}. This procedure reduce the number of $\xi^{\prime}$s and we end up with a Lagrangian which has the structure given in \rf{lag1}. Then the whole procedure can be repeated again until all constraints are eliminated and we are left with a completely reduced, unconstrained and canonical system.

\subsection{The constrained symplectic formalism}

There may be technical difficulties in performing all the steps of the F-J formalism: solving the constraints may prove too difficult, constructing the Darboux transformations may not be possible. An analysis from the point of view of the so called symplectic reduction procedure has been proposed by the authors of \ct{wot, mont1, mont2, barc1}, avoiding the difficult Darboux transformation implied in the original analysis of F-J by expanding, at each stage of the algorithm, the number of variables in the phase space. The central idea in this modified method is to include the constraints into the lagrangian by means of Lagrange multipliers that are velocities, which must be added to the set of dynamical variables, enlarging the configuration space. 
 Once the constraints \rf{const} have been identified, and introduced as strong relations in the symplectic potential $V(\xi)$, one can always rearrange the Lagrangian \rf{lag1} in the form
\be
L^{\prime}\eq a_{i}(q)\dot{q}_{i}+\O^{\a}\dot{\l}_{\a}-V(q),
\ee
where $i=1,...,n$ and $\a =1,...,m$. The coordinates $q_{i}$ describe the non singular sector of $L$, so that the matrix $f_{ij}(q_{i})\eq\frac{\pa}{\pa{q_{i}}^{i}}a_{j}(q_{i})-\frac{\pa}{\pa{q_{i}}^{j}}a_{i}(q_{i})$ is invertible. Then $n$ must be even. 
The new canonical structure is obtained by constructing the matrix $f^{\prime}_{AB}, (A,B =1,...,n, n+1,...,n+m)$ associated to this Lagrangian $L^{\prime}$. If the resulting matrix is nonsingular, the generalized brackets are defined by its inverse. In other case, the whole procedure can be repeated again until the last $f$ matrix be regular. For gauge invariant theories, the algorithm is not able to provide a non singular matrix. To obtain the bracket structure one must include a gauge fixing condition in the symplectic potential as a symmetry breaking term.   

The equivalence of the FJ approach to the Dirac method is discussed in \ct{gov,mont1,garcia}. Some applications of the original FJ and the ``symplectic quantization'' methods can be found in \ct{appli} and the extension to the fermionic case for constant symplectic structures is accomplished in \ct{gov}.

\section{The model}

We are going to study the conformal affine Toda system coupled to matter (Dirac) field (CATM) defined by the two-dimensional field theory  
\br
 {\cal L} = {1\o 4} \pa_{\mu} \vp \, \pa^{\mu} \vp
+ \h  \pa_{\mu} \nu \, \pa^{\mu} \eta
- {1\o 8}\, m_{\psi}^2 \, e^{2\,\eta} 
+ i  {\bar{\psi}} \gamma^{\mu} \pa_{\mu} \psi
- m_{\psi}\,  {\bar{\psi}} \,
e^{\eta+2\vp\,\gamma_5}\, \psi,
\lab{lagrangian}
\er
where ${\bar{\psi}} \equiv {\widetilde \psi}^{T} \,\gamma_0$. This Lagrangian is real for $\eta $ and $\nu$ real fields if $\widetilde{\psi }$ is the
complex conjugate of $\psi$, and if $\varphi $ is pure imaginary.

This model is associated to the principal gradation of the untwisted affine Kac-Moody algebra $sl(2)^{(1)}$, and  belongs to a special
class of models introduced in \ct{matter} called affine Toda systems coupled to matter fields. The zero curvature representation, the construction of the general solution including the solitonic ones and their global and local symmetries were presented in \ct{matter}.

 We are going to mention the relevant symmetries of \rf{lagrangian}. The model \rf{lagrangian} is invariant under the conformal transformations
\be
\lab{conformal}
x_{+} \ra {\hat x}_{+} = f(x_{+}) \, , \qquad 
x_{-} \ra {\hat x}_{-} = g(x_{-}),
\lab{ct}
\ee
with $f$ and $g$ being analytic functions; and with the fields transforming 
as
\br
\nn
\vp (x_{+}\, , \, x_{-}) &\ra& 
{\hat {\vp}}({\hat x}_{+}\, , \,  {\hat x}_{-}) = 
\vp (x_{+}\, , \, x_{-}) \, ,
\\
e^{-\nu (x_{+}\, , \, x_{-})} &\ra& 
e^{-{\hat \nu}({\hat x}_{+}\, , \, 
{\hat x}_{-})} = \( f^{\pr}\)^{\d} \, \( g^{\pr}\)^{\d}
e^{-\nu (x_{+}\, , \, x_{-})} \, ,
\lab{ctf}\\
\nn
e^{-\eta (x_{+}\, , \, x_{-})} &\ra& e^{-{\hat \eta}({\hat x}_{+}\, , \, 
{\hat x}_{-})} = \( f^{\pr}\)^{\h} \, \( g^{\pr}\)^{\h}  e^{-\eta (x_{+}\, 
, \, x_{-})} \, ,
\\ 
\nn
\psi (x_{+}\, , \, x_{-}) &\ra & {\hat {\psi}} ({\hat x}_{+}\, , \, 
{\hat x}_{-}) =   e^{{1\o 2}\( 1+ \gamma_5\) \log \( f^{\pr}\)^{-\h} 
+ {1\o 2}\( 1- \gamma_5\) \log \( g^{\pr}\)^{-\h}}
\, \psi (x_{+}\, , \, x_{-}) \, ,
\er
where the conformal weight $\d$, associated to $e^{-\nu}$, is arbitrary, and 
$\widetilde \psi$ transforms in the same way as $\psi$ (See \ct{matter} for notation).

On the other hand, \rf{lagrangian} is also invariant under the commuting $U(1)_L \otimes U(1)_R$ left and right local gauge transformations
\br
\lab{leri1}
\vp \ra \vp + \xi_{+}\( x_{+}\) + \xi_{-}\( x_{-}\) \; ; \qquad 
\nu \ra \nu \; ; \qquad \eta \ra \eta 
\er
and
\br
\lab{leri2} 
\psi \ra e^{- i\( 1+ \gamma_5\) \xi_{+}\( x_{+}\) 
+ i\( 1- \gamma_5\) \xi_{-}\( x_{-}\)}\, \psi 
\; ; \qquad 
{\widetilde \psi} \ra e^{ i\( 1+ \gamma_5\) \xi_{+}\( x_{+}\) 
- i\( 1- \gamma_5\) \xi_{-}\( x_{-}\)} {\widetilde \psi}. 
\er

By a special choice of $\xi_{+}\( x_{+}\) = - \xi_{-}\( x_{-}\) = - \h
\theta$, with $\theta = {\rm const.}$, one gets a global $U(1)$ transformation 
\be
\vp \ra \vp  \; ; \qquad 
\nu \ra \nu \; ; \qquad \eta \ra \eta   \; ; \qquad 
\psi \ra e^{i \theta} \, \psi  \; ; \qquad 
{\widetilde \psi} \ra e^{-i \theta} \, {\widetilde \psi}, 
\lab{globalu1}
\ee
and the Noether current, associated to the choice, is given by
\be
J^{\mu} = {\bar{\psi}}\, \gamma^{\mu}\, \psi \, , \qquad
\pa_{\mu}\, J^{\mu} = 0.
\lab{noethersl2}
\ee

Another choice could also be possible by taking $\xi_{+}\( x_{+}\) = \xi_{-}\( x_{-}\) = - \h\alpha$, with $\alpha = {\rm const.}$. In this way, one has the global chiral symmetry, too
\be
\psi \ra e^{i\gamma_5 \a}\, \psi \; ; \qquad 
{\widetilde \psi}\ra e^{-i\gamma_5 \a}\, {\widetilde \psi} \; ; \qquad 
\vp \ra \vp -  \a \; ; \qquad 
\nu \ra \nu \; ; \qquad \eta \ra \eta,
\ee
and the corresponding Noether current is 
\be
J_5^{\mu} =  \bar \psi \gamma^\mu \gamma_5 \psi 
+{1\over 2} \partial^\mu \vp  
\; ; \qquad \qquad \pa_{\mu}J_5^{\mu} =0. 
\lab{chiral}
\ee

Concerning the topological current, the Lagrangian \rf{lagrangian} is invariant under $\vp \ra \vp + in \pi$, with all the other fields unchanged. Indeed, the vacua are infinitely degenerate, and the topological charge 
\be
Q_{\rm topol.} \equiv \int \, dx \, j^0 
\, , \qquad
j^{\mu} =  {1\o{2\pi i}}\epsilon^{\mu\nu} \pa_{\nu} \, \vp 
\lab{topological}
\ee
depending only on the asymptotic values of $\vp$, at $x=\pm \infty$, can take non-zero values. 
 
Forward, we are going to discuss the relevance of these transformations which were mentioned former. Associated to the conformal symmetry \rf{conformal} there are two chiral currents 
\be
{\cal J}= - i{\widetilde {\psi}}^T \( 1+\gamma_5 \)\psi 
+ \partial_+\vp +\partial_+\eta, \qquad
{\bar {\cal J}}=  i{\widetilde {\psi}}^T \( 1-\gamma_5 \)\psi 
+ \partial_-\vp+ \partial_-\eta
\lab{chiralcur}
\ee
satisfying 
\be
\partial_-{\cal J}=0 \; ; \qquad \quad  \partial_+{\bar {\cal J}}=0.
\ee

Observe, from \rf{ctf}, that the currents ${\cal J}$ and ${\bar {\cal J}}$ have
conformal weights $(1,0)$ and $(0,1)$ respectively. Under the conformal transformations \rf{conformal}, the chiral currents transform as
\br
\lab{chi1}
\;\;{\cal J}(x_{+})&\longrightarrow& [\ln f^{\pr}( x_{+})
]^{-1}\left({\cal J}(x_{+})-[\ln f^{\pr}(x_{+})]^{\pr}\right),\\
\lab{chi2}
\overline{{\cal J}}(x_{-})&\longrightarrow& [\ln g^{\pr}(x_{-})
]^{-1}\left(\overline{{\cal J}}(x_{-})-[\ln
g^{\pr}( x_{-})]^{\pr}\right)
\er 
 
Then, given a solution of the model, one can always map it, under a conformal transformation, into a solution where

\be
{\cal J}=0 \; ; \qquad \quad {\bar {\cal J}}=0. 
\lab{constraints}
\ee

Such a procedure amounts to ``gauging'' away the free field $\eta$. In other words, \rf{constraints} are constraints implementing a Hamiltonian reduction. This  resembles the connection between the affine (AT) and conformal affine Toda (CAT) models performed in \ct{const, aratyn} through a gauge fixing of the conformal symmetry. The quantum version of the reduction CAT $\rightarrow$ AT was performed in  \ct{bonora}.

Now, we are performing the gauge fixing procedure. Notice from \rf{ctf} that $\vp$ is a scalar under conformal transformations and if we set $\delta$ to zero, $e^{\nu}$ is also scalar. On the other hand $e^{\eta}$ is a $(1/2,1/2)$ primary field. Let us perform a conformal transformation \rf{ctf} with
\br
f^{\pr}(x_{+})=e^{2\eta_+(x_+)},\,\,\,\,g^{\pr}(x_{-})=e^{2\eta_-(x_-)}
\er    
where $\eta_{\pm}(x_{\pm})$ are solutions of the $\eta$ free field ( $\eta(x_+,x_-)=  \eta_+(x_+)+\eta_-(x_-)$, where  $\eta_{\pm}(x_\pm)$ are arbitrary functions), one gets
\br
\hat{\vp} (\hat{x}_{+}\, , \, \hat{x}_{-}) \ra {{\vp}}(x_{+},x_-),\,\,\, e^{-\hat{\eta}_+(\hat{x}_+,\hat{x}_-)}\ra 1,\,\,\, e^{-\hat{\nu}_+(\hat{x}_+,\hat{x}_-)}\ra e^{-\nu_+(x_+,x_-)}. 
\er

Therefore, we are choosing one solution in each orbit of the conformal group. Then for every regular solution of the $\eta$ field the CATM defined on a space-time $(x_+,x_-)$ corresponds to a submodel which we call affine Toda model coupled to matter (ATM) with the extra field $\nu$ defined on a space time $(\hat{x}_+,\hat{x}_-)$. For  the particular solution $\eta = 0$ the CATM and the ATM are defined on the same space-time. Thus setting the $\eta$ field to zero in the equations of motion of the CATM model and reconstructing the Lagrangian for the set of equations of motion of the Toda and Dirac fields we may obtain
\br
\lab{lagrange}
\frac{1}{k}{\cal L}(\varphi ,\psi ,\overline{\psi })=\frac{1}{4}\partial_{\mu }\varphi \partial ^{\mu }\varphi +i\overline{\psi }\gamma ^{\mu
}\partial _{\mu }\psi -m_{\psi }\overline{\psi }e^{2\varphi
\gamma _{5}}\psi. 
\er   

This Lagrangian is not conformal invariant and defines the ATM model.     
One can easily check that the constraints \rf{constraints}, once we have set $\eta=0$, are equivalent to 
\be
{1\o{2\pi i}}\epsilon^{\mu\nu} \pa_{\nu} \, \vp=
{1\o \pi} \bar \psi \gamma^\mu  \psi.
\lab{equivcurrents}
\ee
Thus, in the gauge fixed model, the Noether  current \rf{noethersl2} is
proportional to the topological current \rf{topological}. For
instance, it implies  that the charge density $\psi^{\dagger}\psi$ is
proportional to the space derivative of $\vp$; consequently, the matter field
get confined inside the solitons. It can be seen in the case of one-soliton and two-soliton solutions of \rf{lagrangian} which was calculated in \ct{matter,bla}.

The vaccum solution $\eta=0$ was used in \ct{matter,bla} to perform the dressing transformation in order to obtain soliton solutions, which are in the orbit of a vacuum solution. It can be also noticed that these transformations do not excite the field $\eta$ and the solitonic solutions are solutions of the gauge fixed model. 
The one-soliton solution for $\vp$ is exactly a sine-Gordon type soliton, and the corresponding $\psi$ solution is of the massive Thirring model type \ct{orfa}. In addition, one can check that these solutions satisfy \rf{equivcurrents}, and so it is a solution of the gauge fixed model, implying that the Dirac field must be confined inside the solitons. 

\section{The FJ formalism and the massive Thirring model}

As it was discussed in the last section we will consider the following Lagrangian
\be
\lab{lagrangian1}
\frac{1}{k}{\cal L}(\varphi ,\psi ,\overline{\psi },\lambda_{\mu})=\frac{1}{4}\partial_{\mu }\varphi \partial ^{\mu }\varphi +i\overline{\psi }\gamma ^{\mu
}\partial _{\mu }\psi -m_{\psi }\overline{\psi }e^{2\varphi
\gamma _{5}}\psi +\lambda_{\mu}(2i\overline{\psi}\gamma^{\mu}\psi-\epsilon^{\mu\nu}\partial_{\nu}\varphi), 
\ee
where we have incorporated the Noether and Topological currents equivalence \rf{equivcurrents} by adding to the Lagangian the term $\lambda_{\mu}(2\overline{\psi}\gamma^{\mu}\psi-\epsilon^{\mu\nu}\partial_{\nu}\varphi)$, where $\lambda_{\mu}$ are Lagrange multipliers. The same procedure has been used, for example, to incorporate the left-moving condition for the boson in the study of chiral bosons in two dimensions \ct{siegel}. Notice that the $\lambda_{\mu}$ terms in \rf{lagrangian1} will break the left-right local symmetries \rf{leri1}-\rf{leri2} of ATM \rf{lagrange} which survived the conformal gauge fixing. 
In order to write \rf{lagrangian1} in the first order form \rf{lag1}, let us define the conjugated momenta 
\be
\lab{moments}
\pi _{\varphi}=\frac{1}{2}\dot{\varphi}+\lambda_{1},\qquad \pi _{\lambda_{\mu}}=0,\qquad \pi _{\psi_{R}}=-i\psi_{R},\qquad \pi _{{\psi}_{L}}=-i\psi_{L}.
\ee

We are assuming that the Dirac fields are anticommuting Grasmannian variables and their momenta variables defined through {\bf left} derivatives.   
Then, as usual, the Hamiltonian is defined by

\be
\lab{hamiltonian}
{\cal H}_{c}=\pi _{\varphi}\dot{\varphi}+\pi _{{\psi}_{R}}\dot{\psi_{R}}+\pi _{{\psi}_{L}}\dot{\psi}_{L}-\cal{L}.
\ee
Explicitely the Hamiltonian density becomes 
\br
\lab{hamiltonian1}
{\cal H}_{c}&=&\pi _{\varphi}^{2}+ \lambda_{1}^2+\frac{1}{4}{\varphi^{\prime}}^{2}+\pi _{{\psi}_{R}}\psi_{R}^{\prime}-\pi _{{\psi}_{L}}\psi_{L}^{\prime}-2i\lambda_{1}j^{1}-2\lambda_{1}\pi_{\varphi}-\lambda_{0}(2ij^{0}-\varphi^{\prime})+\\
\nn
& &im_{\psi}(e^{-2\varphi}\widetilde{\psi}_{R}\psi_{L}-e^{2\varphi}\widetilde{\psi}_{L}\psi_{R}). 
\er
Now, the same Legendre transform \rf{hamiltonian} is used to write the first order Lagrangian 
\br
\lab{lagran}
{\cal L}&=\pi _{\varphi}\dot{\varphi}+\pi _{{\psi}_{R}}\dot{\psi_{R}}+\pi _{{\psi}_{L}}\dot{\psi}_{L}-{\cal H}_{c}.
\er

Our starting point for the F-J analysis will be this first order Lagrangian. The Lagrangian \rf{lagran} is already in the form \rf{lag1}, and the Euler-Lagrange equations for the Lagrange multipliers allow us to solve one of them 
\be
\lambda_{1}=ij^{1}+\pi_{\varphi},
\ee
and the other equation leads to one constraint
\be
\lab{cons}
\O_{1}\equiv2ij^{0}-\varphi^{\prime}=0.
\ee

Then, the Lagrange multiplier $\lambda_{1}$ must be replaced back in \rf{lagran} and the constraint \rf{cons} solved. Firstly, let us replace the $\lambda_{1}$ multiplier into ${\cal H}_{c}$, then one gets
\br
\lab{hamiltonian2}
{\cal H}_{c}^{\prime}&=&(j^{1})^{2}+\frac{1}{4}{\varphi^{\prime}}^{2}  +\pi _{{\psi}_{R}}\psi_{R}^{\prime}-\pi _{{\psi}_{L}}\psi_{L}^{\prime} -2i\pi_{\varphi}j^{1}-\lambda_{0}\O_{1}+\\
\nn
& &im_{\psi}e^{-2\varphi}\widetilde{\psi}_{R}\psi_{L}-e^{2\varphi}\widetilde{\psi}_{L}\psi_{R}). 
\er
Then the new Lagrangian becomes
\be
\lab{lagran1}
{\cal L}^{\prime}=\pi _{\varphi}\dot{\varphi}+\pi _{{\psi}_{R}}\dot{\psi_{R}}+\pi _{{\psi}_{L}}\dot{\psi}_{L}-{\cal H}_{c}^{\prime}.
\ee
One can implement the constraint by replacing in this Lagrangian the field $\varphi$ in terms of the space integral of the current component $j^{0}$. Notice that in the interaction terms there will arise a complicated non-local expressions. Thus we get the following Lagrangian 
\begin{eqnarray}
\label{lagran2}
{\cal L}^{\prime\prime}&=&\pi_{\varphi}\pa_{t}{\int^{x}2ij^{0}} + 
	\pi _{{\psi}_{R}}\dot{\psi_{R}} + \pi _{{\psi}_{L}}\dot{\psi}_{L} 
	+ i\widetilde{\psi}_{R}\psi_{R}^{\prime} 
	- i\widetilde{\psi}_{L}{\psi}_{L}^{\prime} 
	- \\
\nonumber
& & im_{\psi}(e^{-4i\int^{x}j^{0}}\widetilde{\psi}_{R}
	\psi_{L}
 - e^{4i\int^{x}j^{0}}\widetilde{\psi}_{L}\psi_{L}) 
	+ j^{\mu}j_{\mu}+ 2i\pi_{\varphi}j^{1}.
\end{eqnarray}
Notice the appearence of the current-current interaction term. The following Darboux transformations
\be
\lab{darboux}
\pi_{\varphi} \rightarrow  \pi_{\varphi}^{\prime},\qquad 
\psi_{R} \rightarrow e^{-2i\int^{x}j^{0}}\psi_{R}, \qquad  \psi_{L} \rightarrow e^{2i\int^{x}j^{0}}\psi_{L},
\ee
with the $\pi_{\varphi}^{\prime}$ momenta convenientely chosen, is used to diagonalize the canonical one-form. We are, thus, after defining $k\equiv-2/g$, and rescaling the field $\psi \rightarrow 1/\sqrt{k} \psi$, left with the Lagrangian 
\begin{equation}
\lab{thirring1}
{\cal L}[\psi,\overline{\psi}]= i\overline{\psi}\gamma^{\mu}\pa_{\mu}\psi-m_{\psi}\overline{\psi}\psi-\frac{1}{2}g j_{\mu}j^{\mu}
\end{equation}
which is just the massive Thirring model. The canonical pairs are  $(-i\widetilde{\psi}_{R}, \psi_{R})$ and $(-i\widetilde{\psi}_{L}, \psi_{L})$.   

\section{The constrained symplectic formalism and the massive Thirring system}

Now let us study our model in the framework of the symplectic formalism. The superscript numbers will be written in accordance to the iterative character of the constrained symplectic formalism. Thus, ${\cal L}^{\prime}$ is considered as the zeroth-iterated Lagrangian ${\cal L}^{(0)}$. In order to perform this analysis we will start from the first order lagrangian \rf{lagran1}, written as 
\be
{\cal L}^{(1)}=\pi_{\varphi}\dot{\varphi}+\pi_{\psi_{R}}\dot{\psi}_{R}+\pi_{\psi_{L}}\dot{\psi}_{L}+\dot{\eta}^{1}\O_{1}-{\cal V}^{(1)},
\ee
where the once-iterated symplectic potential is
\be
{\cal V}^{(1)}={\cal H}_{c}^{\prime}|_{\Omega_{1}=0}, 
\ee
or explicitly
\br
{\cal V}^{(1)}&=&[(j^{1})^{2}-2i\pi_{\varphi}j^{1}+\frac{1}{4}{\varphi^{\prime}}^{2}-i\widetilde{\psi}_{R}\psi_{R}^{\prime}+i\widetilde{\psi}_{L}\psi_{L}^{\prime}+\\
\nonumber
& & im_{\psi}(e^{-2\varphi}\widetilde{\psi}_{R}\psi_{L}-e^{2\varphi}\widetilde{\psi}_{L}\psi_{R})]|_{\O_{1}=0} 
\er
and the stability condition of the symplectic constraint $\O_{1}$ under time evolution has been implemented by doing $\l_{0} \rightarrow \dot{\eta}^{1}$. Consider the once-iterated set of symplectic variables in the following order 
\be
\xi^{(1)}_{A}=(\eta^{1},\varphi,\psi_{R},\psi_{L},\pi_{\varphi},\pi_{\psi_{R}},\pi_{\psi_{L}})
\ee
and the components of the canonical one-form
\be
a_{A}^{(1)}=(\O_{1},\pi_{\vp},\pi_{\psi_{R}},\pi_{\psi_{L}},0,0,0).
\ee

These result in the following singular symplectic two-form matrix
\br
f^{(1)}_{AB}(x,y)&=&\left(\begin{array}{ccccccc}0 & -\pa_{x} & -2i\widetilde{\psi}_{R} & -2i\widetilde{\psi}_{L} & 0 & -2\psi_{R} & -2\psi_{L}\\
-\pa_{x} & 0 & 0 & 0 & -1 & 0 & 0\\
-2i\widetilde{\psi}_{R} &  0 & 0 & 0 & 0 & -1 & 0\\
-2i\widetilde{\psi}_{L} & 0 & 0 & 0 & 0 & 0 & -1\\
0 & 1 & 0 & 0 & 0 & 0 & 0\\
-2\psi_{R} & 0 & -1 & 0 & 0 & 0 & 0\\
-2\psi_{L}& 0 & 0 & -1 & 0 & 0 & 0 
\end{array}\right)\d(x-y).
\er
This matrix has a zero mode
\be
{\bf {{v}}^{(1)}}^{T}(x)=\left(u,0,-2u\psi_{R},-2u\psi_{L},-u^{\prime},-2iu\widetilde{\psi}_{R},-2iu\widetilde{\psi}_{L}\right),
\ee
where $u$ is an arbitrary function. The zero-mode condition, Eq. \rf{const} gives
\be
\int {dx {\bf{{v}}^{(1)}}^T(x)\frac{\d}{\d \Phi(x)}\int{dy{\cal V}^{(1)}}}\equiv0.
\ee
Then the gradient of the symplectic potential happens to be orthogonal to the zero-mode ${\bf {v}}^{(1)}$. The equations of motion are automatically validated, and no symplectic constraint appear. This happens due to the presence of the following symmetry of the action  
\begin{eqnarray}
\d\eta^{1}&=&u,  \d\varphi=0, \d\pi_{\varphi}=-u^{\prime},\;\;\d\psi_{R}=
	-2u\psi_{R}, \;\;\d\psi_{L}=-2u\psi_{L},\\
\nonumber
\d\pi_{\psi_{R}}&=&-2iu\widetilde{\psi}_{R}, \;\;\d\pi_{\psi_{L}}=-2iu\pi_{\psi_{L}}.
\end{eqnarray}
So, in order to deform the symplectic matrix into an invertible one, we have to add a gauge fixing term to the symplectic potential. One can choose any consitent gauge fixing condition \ct{mont1}. In this case we have one symmetry generator associated to the parameter $u$, so there must be one gauge condition. Let us choose the gauge
\be
\lab{gauge1}
\O_{2}\equiv\vp=0.
\ee
By doing this we are gauging away the $\vp$ field, so only the remaining field variables will describe the dynamics of the system. Other gauge conditions, which gauge away the $\psi$ field will be imposed in the next subsection.  Implementing the consistency condition by means of Lagrange multiplier $\eta^{2}$ (which enlarges even further the configuration pace) we get the twice-iterated Lagrangian
\be
{\cal L}^{(2)}=\pi_{\varphi}\dot{\varphi}+\pi_{\psi_{R}}\dot{\psi}_{R}+\pi_{\psi_{L}}\dot{\psi}_{L}+\dot{\eta}^{1}\Omega_{1}+\dot{\eta}^{2}\Omega_{2}-{\cal V}^{(2)},
\ee
where
\br
{\cal V}^{(2)}\;&=&\;{\cal V}^{(1)}|_{\Omega_{2}=0}\\
\nn
&=&-2ij^{1}\pi_{\vp}-i\widetilde{\psi}_{R}\psi^{\prime}_{R}+i\widetilde{\psi}_{L}\psi^{\prime}_{L}+m_{\psi}e^{\eta_{0}}\overline{\psi}\psi-j^{\mu}j_{\mu}
\er
or explicitly
\be
{\cal V}^{(2)}=-j^{\mu}j_{\mu}-2ij^{1}\pi_{\vp}-i\widetilde{\psi}_{R}\psi_{R}^{\prime}+i\widetilde{\psi}_{L}\psi_{L}^{\prime}+m_{\psi}\overline{\psi}\psi.
\ee

Assuming now that the new set of symplectic variables is given in the following order 
\be
\xi^{(2)}_{A}=(\eta^{1},\eta^{2},\varphi,\psi_{R},\psi_{L},\pi_{\varphi},\pi_{\psi_{R}},\pi_{\psi_{L}})
\ee
and the non vanishing components of the canonical one-form
\be
a_{A}^{(1)}=(\O_{1},\O_{2},\pi_{\vp},\pi_{\psi_{R}},\pi_{\psi_{L}}, 0, 0, 0),
\ee
we can obtain the singular twice-iterated symplectic matrix 
\br
f^{(2)}_{AB}(x,y)&=&\left(\begin{array}{cccccccc}0 & 0 & -\pa_{x} & -2i\widetilde{\psi}_{R} & -2i\widetilde{\psi}_{L} & 0 & -2\psi_{R} & -2\psi_{L}\\
0 & 0 & -1 & 0 & 0 & 0 & 0 & 0\\
-\pa_{x} & 1 & 0 & 0 & 0 & -1 & 0 & 0\\
-2i\widetilde{\psi}_{R} &  0 & 0 & 0 & 0 & 0 & -1 & 0\\
-2i\widetilde{\psi}_{L} & 0 & 0 & 0 & 0 & 0 & 0 & -1\\
0 & 0 & 1 & 0 & 0 & 0 & 0 & 0\\
-2\psi_{R} & 0 & 0 & -1 & 0 & 0 & 0 & 0\\
-2\psi_{L} & 0 & 0 & 0 & -1 & 0 & 0 & 0 
\end{array}\right)\d(x-y).
\er
The corresponding zero-mode is  
\be
{\bf {{v}}^{(2)}}^T(x)=\left(u, 0, 0, -2u\psi_{R}, -2u\psi_{L}, 0, -2iu\widetilde{\psi}_{R}, -2iu\widetilde{\psi}_{L}\right)
\ee
and the following symmetry of the action
\br
\d\eta^{1}&=&u,\d\eta^{2}=0,\;\;\d\varphi=0,\;\;\d\pi_{\varphi}=0,\;\;\d\psi_{R}=-2u\psi_{R},\\
\nn
\;\;\d\psi_{L}&=&-2u\psi_{L},\;\;\d\pi_{\psi_{R}}=-2iu\widetilde{\psi}_{R},\;\;\d\pi_{\psi_{L}}=-2iu\pi_{\psi_{L}}.
\er
This time let us choose the gauge
\be
\lab{gauge2}
\O_{3}\equiv\pi_{\vp}=0,
\ee
and using  the consistency condition by means of a new Lagrange multiplier $\eta^{3}$ we get the third-iterated Lagrangian
\be
{\cal L}^{(3)}=\pi_{\varphi}\dot{\varphi}+\pi_{\psi_{R}}\dot{\psi}_{R}+\pi_{\psi_{L}}\dot{\psi}_{L}+\dot{\eta}^{1}\Omega_{1}+\dot{\eta}^{2}\Omega_{2}+\dot{\eta}^{3}\Omega_{3}-{\cal V}^{(3)},
\ee
where
\be
{\cal V}^{(3)}={\cal V}^{(2)}|_{\Omega_{3}=0}.  
\ee
or explicitly
\be
{\cal V}^{(3)}=-j^{\mu}j_{\mu}-i\widetilde{\psi}_{R}\psi_{R}^{\prime}+i\widetilde{\psi}_{L}\psi_{L}^{\prime}+m_{\psi}\overline{\psi}\psi.
\ee
Since the symplectic two-form for this Lagrangian happens to be a non singular matrix our algorithm has come to an end. 

Now it is easy to see that collecting the canonical part and the symplectic potential ${\cal V}^{(3)}$ we end up with the following massive Thirring model Lagrangian

\be
\lab{thirring2}
{\cal L}[\psi,\overline{\psi}]= i\overline{\psi}\gamma^{\mu}\pa_{\mu}\psi-m_{\psi}\overline{\psi}\psi-\frac{1}{2}g j_{\mu}j^{\mu}+\mu j^{0}. 
\ee

We have made the same choice, $k=-2/g$, as in the previous section, and the field rescaling $\psi \rightarrow 1/\sqrt{k} \psi$. This is in agreement with our result in \rf{thirring1}, where it has been obtained solving the constraint and diagonalizing the canonical part by Darboux transforming the Dirac field \rf{darboux}. As a bonus, in the symplectic approach we get a term associated to the chemical potential $\mu$ ( $-i\dot{\eta}^{1} \rightarrow \mu $) times the fermion charge density.  

\subsection{The sine-Gordon model}

As it was anticipated above, one can choose another gauge fixing, instead of \rf{gauge1}, to construct the twice-iterated Lagrangian. Let us make the choice
\be
\lab{gauge3}
\O_{2}\equiv j^{0}=0,
\ee
which satisfies the non-gauge invariance condition as can easily be verified by computing the bracket $\{\O_{1}\, , \, j^{0}\} =0$. The twice-iterated Lagrangian is obtained by bringing back this constraint into the canonical part of ${\cal L}^{(1)}$, then

\be
{\cal L}^{(2)}=\pi_{\varphi}\dot{\varphi}+\pi_{\psi_{R}}\dot{\psi}_{R}+\pi_{\psi_{L}}\dot{\psi}_{L}+\dot{\eta}^{1}\O_{1}+\dot{\eta}^{2}\O_{2}-{\cal V}^{(2)},
\ee
where the twice-iterated symplectic potential becomes
\be
{\cal V}^{(2)}={\cal V}^{(1)}|_{\Omega^{2}=0}. 
\ee

Considering the set of symplectic variables in the following order
\be
\xi^{(2)}_{A}=(\eta^{1},\eta^{2},\varphi,\psi_{R},\psi_{L},\pi_{\varphi},\pi_{\psi_{R}},\pi_{\psi_{L}})
\ee
and the components of the canonical one-form
\be
a_{A}^{(2)}=(\O_{1},\O_{2},\pi_{\vp},\pi_{\psi_{R}},\pi_{\psi_{L}},0,0,0)
\ee
the (degenerated) symplectic matrix is easily found to be
\br
f^{(2)}_{AB}(x,y)=\left(\begin{array}{cccccccc}0 & 0 & -\pa_{x} & -2i\widetilde{\psi}_{R} & -2i\widetilde{\psi}_{L} & 0 & -2\widetilde{\psi}_{R} & -2\widetilde{\psi}_{L}\\
0 & 0 & 0 & -\psi_{R} & -\psi_{L} & 0 & i\psi_{R} & i\psi_{L}\\
-\pa_{x} & 0 & 0 & 0 & 0 & -1 & 0 & 0\\
-2i\widetilde{\psi}_{R} & -\widetilde{\psi}_{R} & 0 & 0 & 0 & 0 & -1 & 0\\
-2i\widetilde{\psi}_{L} & -\widetilde{\psi}_{L} & 0 & 0 & 0 & 0 & 0 & -1\\
0 & 0 & 1 & 0 & 0 & 0 & 0 & 0\\
-2\psi_{R} & i\psi_{R} & 0 & -1 & 0 & 0 & 0 & 0\\
-2\psi_{L} & i\psi_{L} & 0 & 0 & -1 & 0 & 0 & 0\\
\end{array}\right).\\
\nn
\d(x-y).
\er
Its zero-mode is easily found to be
\be
{\bf {{v}}^{(2)}}^T(x)=(u,v,0,(iv-2u)\psi_{R},(iv-2u)\psi_{L},-u^{\prime},(-v-2iu)\widetilde{\psi}_{R},(-v-2iu)\widetilde{\psi}_{L}),
\ee
where $u$ and $v$ are arbitrary functions. The zero mode condition, Eq. \rf{const} selects the following constraint
\be
\int {dx {\bf {v}}^{(2))^{T}}(x)\frac{\d}{\d \Phi(x)}\int{dy{\cal V}^{(2)}}}=\int{dx j^{1}\partial_{x}v}\equiv0.
\ee
Since the function $\partial_{x}v$ is arbitrary we end up with the following Lagrangian constraint
\be
\lab{lagcons}
\O_{3}\equiv j^{1}=0.
\ee 
Notice that by solving the constraints $\O_{2}=0$ and $\O_{3}=0$, we obtain
\be
\lab{majorana}
\widetilde{\psi}_{R} = \psi_{R},\;\;\;\widetilde{\psi}_{L} = \psi_{L}.
\ee 
This is precisely the reality conditions for the Fermi fields. So, we end up in the theory with a Majorana fermion, the scalar $\vp$, and the auxiliary fields. 
With the consistency condition implemented by introducing a third Lagrange multiplier into ${\cal L}^{(2)}$ we get
\be
{\cal L}^{(3)}=\pi_{\varphi}\dot{\varphi}+\pi_{\psi_{R}}\dot{\psi}_{R}+\pi_{\psi_{L}}\dot{\psi}_{L}+\dot{\eta}^{1}\O_{1}+\dot{\eta}^{2}\O_{2}+\dot{\eta}^{3}\O_{3}-{\cal V}^{(3)},
\ee
where
\be
{\cal V}^{(3)}={\cal V}^{(2)}|_{\Omega_{3}=0}
\ee
or
\be
{\cal V}^{(3)}={1\over 4}{\vp^{\prime}}^2-i{\psi}_{R}\psi_{R}^{\prime}+{\psi}_{L}\psi_{L}^{\prime}+im_{\psi}{\psi}_{R}\psi_{L} (e^{-2\vp }+ e^{2\vp}).
\ee

The new set of symplectic variables is assumed to be ordered as
\be
\xi^{(3)}_{A}=(\eta^{1},\eta^{2},\eta^{3}, \varphi, \psi_{R},\psi_{L},\pi_{\varphi},\pi_{\psi_{R}},\pi_{\psi_{L}}).
\ee
By inspection we find the following components of the canonical one-form
\be
a_{A}^{(3)}=(\O_{1},\O_{2},\O_{3},\pi_{\vp},\pi_{\psi_{R}},\pi_{\psi_{L}},0,0,0).
\ee

After some algebraic manipulations we get the following third-iterated symplectic two-form
\br
f^{(3)}_{AB}(x,y)=\left(\begin{array}{ccccccccc}0 & 0 & 0 & -\pa_{x} & -2i\widetilde{\psi}_{R} & -2i\widetilde{\psi}_{L} & 0 & -2\psi_{R} & -2\psi_{L}\\
0 & 0 & 0 & 0 & -\widetilde{\psi}_{R} & -\widetilde{\psi}_{L} & 0 & i{\psi}_{R} & i{\psi}_{L}\\
0 & 0 & 0 & 0 & -\widetilde{\psi}_{R} & \widetilde{\psi}_{L} & 0 & i{\psi}_{R} & -i{\psi}_{L}\\
-\pa_{x} & 0 & 0 & 0 & 0 & 0 & -1 & 0 & 0\\
-2i\widetilde{\psi}_{R} & -\widetilde{\psi}_{R} & -\widetilde{\psi}_{R} & 0 & 0 & 0 & 0 & -1 & 0\\
-2i\widetilde{\psi}_{L} & -\widetilde{\psi}_{L} & \widetilde{\psi}_{L} & 0 & 0 & 0 & 0 & 0 & -1 \\
0 & 0 & 0 & 1 & 0 & 0 & 0 & 0 & 0\\
-2\psi_{R} & i\psi_{R} & i\psi_{R} & 0 & -1 & 0 & 0 & 0 & 0\\
-2\psi_{L} & i\psi_{L} & -i\psi_{L} & 0 & 0 & -1 & 0 & 0 & 0\\
\end{array}\right).\\
\nn
.\d(x-y)
\er
It can be checked that this matrix has the unique zero-mode
\be
{\bf {v}}^{(3)}(x)=(u, -2iu, 0, 0, 0, 0, -u^{\prime}, 0, 0),
\ee
where the function $u$ is totally arbitrary. The zero-mode condition \rf{const} gives rise to
\be
\int {dx {\bf {v}}(x)^{T}\frac{\d}{\d \Phi(x)}\int dy{\cal V}^{(3)}}\equiv0.  
\ee
Then the action has the following symmetry
\br
\d\eta^{1}=u,&\d\eta^{2}=-2iu,& \d\eta^{3}=0,\;\;\d\varphi=0,\;\;\d\pi_{\varphi}=-u^{\prime},\;\;\d\psi_{R}=0,\;\;\d\psi_{L}=0\\
\nn
\d\pi_{\psi_{R}}=0,&\d\pi_{\psi_{L}}=0.
\er

This symmetry allows us to fix $i\psi_{R}\psi_{L}$ to be a constant. By taking $\psi_{R}=-iM \overline{\theta}$ and $\psi_{L}=\theta$, with $M$ a real number, we find that $i\psi_{R}\psi_{L}$ indead becomes a constant. Note that $\theta$ and $\overline{\theta}$ are Grassmannian variables, while $\overline{\theta}\theta$ is an ordinary commuting number. In order to obtain a manifestly Lorentz invariant final Lagrangian we must choose the conjugated momenta $\pi_{\vp}=-\dot{\vp}/4$. Choosing $k=-2/g$ as the overall coupling constant and rescaling the field, $\vp \rightarrow \sqrt{g} \vp$, we are left with 
\be
\lab{sine}
{\cal L^{''}}=\frac{1}{2}(\partial_{\mu}\varphi)^{2}+\frac{2}{g}m_{\psi}M\; \mbox{cos}2\sqrt{g}\varphi+\mu\varphi^{\prime}.
\ee
   
This is just the sine-Gordon model Lagrangian. In addition we get a term multiplied by a chemical potential $\mu$ ($\dot{\eta}^{1} \rightarrow \mu$). This is a topological term, in the sense that it only depends on the value of the $\vp$ field at the spatial boundary ($x= \pm\infty$) which is just the topological charge density.   

\section{Summary and conclusions}

The conformal invariance of the affine $sl_{2}$ Toda model coupled to matter field (CATM) is gauge fixed setting $\eta=$0, and then we have defined de ATM model. Then the reduction is performed by imposing a constraint which is precisely the equivalence relation between the Noether and topological currents. We incorporate this constraint by adding to the action the term $\lambda_{\mu}(2ij^{\mu}-\epsilon^{\mu\nu}\partial_{\nu}\vp)$ where $\l_{\mu}$ are lagrange multipliers. We write the Lagrangian density as a first order expression in time derivatives of the dynamical variables, and observe that $\l_{1}$ appears non linearly while $\l_{0}$ does linearly. The Euler-Lagrange equations for the $\l$'s allow one to solve for the $\l_{1}$ in terms of the other variables, and leave for further analysis the linearly occuring $\l_{0}$ which is related to a true constraint $\O_{1}\equiv 2ij^{0}-\vp^{\prime}$. Then following the initial FJ method we solved the constraint for the $\vp$ field, which is the easiest way to do it, replaced back into the Lagrangian density and performed Darboux's tranformations, and ended up with a canonical expression for the Dirac fields, which combined to the symplectic potential was precisely the massive Thirring model.\\
The second main point we should highlight relies upon the use of the constrained symplectic formalism, which allows to overcome the difficult task of solving the constraint for the Dirac field in terms of the $\vp$ field as required in the initial FJ approach. By bringing the constraints, iteratively into the canonical part of the Lagrangian, considering the zero-mode conditions and after imposing the new constraints and gauge fixing we have been able to arrive finally at the sine-Gordon or massive Thirring model, depending on the specific gauge fixing.\\
An interesting observation is that the method allowed us to obtain the canonical structure only in terms of the fermion or the boson degree of freedom, by conveniently choosing the constraint in the initial coupled Lagrangian \rf{lagrangian1}, which is nothing but the Noether and topological current equivalence condition incorporated making use of Lagrange multipliers, and the convenient choice of gauge fixings Eqs. \rf{gauge1} and \rf{gauge2} to obtain the massive Thirring model, and the gauge fixing Eq. \rf{gauge3} and the Lagrangian constraint Eq.\rf{lagcons} to get the sine-Gordon model. Thus, Eq. \rf{lagrangian1} defines a so called parent Lagrangian \ct{hje} from which both of the models are derivable. From the Lagrangians \rf{thirring1} and \rf{sine}, it is also clear the duality exchange of the coupling regimes: $g \rightarrow 1/g$. Moreover, we believe that this approach clarifies the role played by such parent Lagrangians. For example in \ct{bogoje}, the authors have recently shown the exact quantum equivalence between the sine-Gordon and massive Thirring models by gauge fixing a wider gauge invariant theory in two different ways.    
\section{Acknowledgements}
HB is thankful to FAPESP (grant number 96/00212-0) and BMP to CNPq for financial support.

\end{document}